# ON THE PHASE GROUPING MECHANISM FOR A MAGNETRON COHERENT OSCILLATION

G. Kazakevich#, R.P. Johnson, Muons, Inc, Batavia, IL 60510, USA
V. Yakovlev, Fermilab, Batavia, IL 60510, USA

*Abstract*

CW magnetrons, designed and optimized for industrial heaters, driven by an injection-locking signal, i.e., operating as forced oscillators, were suggested in number of works to power Superconducting RF (SRF) cavities due to higher efficiency and significantly lower cost of generated RF power per Watt than traditionally used RF sources (klystrons, IOTs, solid-state amplifiers). When the magnetrons are intended to feed Room Temperature (RT) cavities, the injected phase/frequency locking signal may provide required phase/frequency stability of the accelerating field. However, when the magnetron RF sources are intended to feed high Q-factor SRF cavities, the sources must be controlled in phase and power in a wide bandwidth to compensate parasitic phase and amplitude modulations caused by "microphonics". In dependence on parameters of magnetron and the injection-locking signal one can choose regime most suitable for feeding SRF cavities, enabling magnetron almost coherent oscillation at the wide bandwidth of control. The paper considers impact of the phase grouping on a quasi-coherent oscillation of magnetrons indicating novel features and capabilities of the RF sources for modern superconducting accelerators.

## INTRODUCTION

Phase/frequency stabilization of a magnetron with an injected RF signal by a frequency pulling in a magnetron due to a mismatched load was considered in [1]. This method allowed stabilizing the accelerating field in the cavity of a microtron-injector of a Terahertz FEL, with a magnetron RF source, by the wave reflected from the accelerating cavity [2]. This work demonstrated first lasing in a FEL driven by a microtron. The magnetron stabilization was simulated by abridged equations of the coupled system of magnetron-accelerating cavity and was verified by experiments [3]. A method stabilizing accelerating field in the cavity powered by a magnetron with an external locking signal was proposed in [4].

For a 2.45 GHz SRF cavity operating at 2 K the phase control of a magnetron by a phase-modulated injection-locking signal was demonstrated in Ref. [5]. The methods for control of the magnetron phase and power in a wide bandwidth have been developed and described in [6-8].

All methods use a phase-modulated injection-locking signal for the phase control. The wide bandwidth vector power control was realized with vector summing by a 3 dB hybrid of output signals of a controlled in phase, two-channel magnetron RF source [6], or in a single-channel RF source controlling the spectral power at the carrier frequency by modulation of depth of the phase modulation [7]. If the frequency modulating the depth is much larger than the SRF cavity bandwidth, the sidebands caused by the modulation are reflected from the SRF cavity into a dummy load. Both vector methods of power control use fast redistribution of the RF power between the RF cavity and a dummy load; this reduces the average efficiency at the vector power control.

The method of power control described in [8] uses a magnetron current control in a tube driven by a large injected signal (≈ -10 dB of the tube nominal power). In this case the magnetron stably generates low power at low current with the anode voltage below the self-excitation threshold. This method provides ≈10 dB range of power control with average efficiency significantly higher compared to the mentioned above vector methods and with the spectral density of noise power lower by ≈20 dB/Hz than provides a magnetron injection-locked by a small signal (of $\leq$ -20 dB) [8, 9]. Combining this method with one of vector methods by Low Level RF (LLRF) techniques enables a wideband phase and power control of magnetrons with highest efficiency.

A large injection-locking signal at the anode voltage below the self-excitation threshold enables a better phase grouping of charges delivered onto the magnetron anode and a reduction of fluctuations of self-consistent field of the synchronous wave [9]. This causes higher efficiency [10], and lower noise in operation of magnetrons. Presented developed parameterization of the phase grouping in a magnetron RF generation demonstrates capabilities to find methods of operation and control of magnetrons that are most suitable for SRF accelerators.

## PARAMETERIZATION OF A PHASE GROUPING IN MAGNETRONS

In a kinetic model [9], we demonstrated the principle of operation of a magnetron basing on a resonant energy exchange between the synchronous wave, rotating in the magnetron interaction space, and the phase-grouped Larmor electrons moving in "spokes" towards the anode. A simplified parameterization of the phase grouping in an injection-locked magnetron is considered below.

Without phase grouping magnetrons do not operate, but some spontaneous oscillations exist in RF system of the injection-locked magnetron even fed below the voltage of start-up.

Estimations show that the RF oscillations in the space of interaction with intensity above the critical, that is necessary for a magnetron launching, one can hardly excite due to fluctuations of charges in an electron cloud blanketing the cathode [11]. However, the sources of RF oscillation can be non-stationary oscillations of alternating quasi-static fields in the electron cloud of space charge around the cathode [12]. Also, the cyclotron oscillations of Larmor electrons can be considered as spontaneous oscillations using in the first approximation eq. (1) for cyclotron radiation in a free space [13].

$$P \sim \frac{2e^4 H^2 {v_L}^2}{3m^2 c^5} = \frac{2e^2 \cdot \omega^2 v_L^2}{3c^3} \qquad (1)$$

Here: $H$ is the magnetron magnetic field, $\omega$ is the cyclotron frequency, $m$ is the electron mass, $v_L$ is velocity of a Larmor electron.

For a 2.45 GHz microwave oven magnetron with $H \approx 880$ Oe at the anode voltage of 3.7 kV, $v_L \approx 0.097c$, $P$ value is $\sim 1.14 \cdot 10^{-18}$ W. Then, total power of the phase locked cyclotron oscillations, considering variation of $v_L$ along a Larmor revolution is ~1 W at the emitted current of 1 A.

For both noted sources of spontaneous oscillations the injection-locking signal destroys their uniform phase distribution providing some phase locking of them. A magnetron RF system stores the phase-locked oscillations as a standing wave. One can represent it in the interaction space as two slow waves with angular velocity $\omega/n$ ($n=N/2$; $N$ is number of vanes) rotating in opposite directions [9].

The slow wave rotating along the Larmor circulation may cause a collective rotation of Larmor orbits with the angular phase velocity of $\omega/n$ due to its azimuthal electric field [9]. Such a wave is a synchronous wave for centers of Larmor orbits on a synchronous radius [12]. The energy required for the collective motion one can estimate using the moment of inertia, $J_e$, of the electron cloud blanketing the cathode as: $W \sim J_e \cdot \omega^2 \cdot p_e / 2n^2$. Here $p_e$ is the relative part of Larmor orbits forming the "spokes". For a microwave oven 2.45 GHz magnetron with $N$=8 the synchronous wave energy $W$~100 J provides a collective rotation of ~1/4 of the electrons that form so called "spokes" in the launched magnetron.

Synchronous motion of electron orbits with the rotating slow wave causes a resonant interaction with the energy exchange of the synchronous wave and moving charges (in the centers of Larmor orbits) [9]. The measurement of phase-locked spontaneous radiation in a magnetron is considered below.

The increasing synchronous wave, upon starting the magnetron, improves the phase synchronism of the spontaneous oscillations of various electrons and groups the electron orbits in the "spokes" in phase. The "spokes" deliver electrons onto the anode, providing magnetron current and RF generation when the azimuthal velocity of moving charges is greater than azimuthal velocity of the rotating wave [9].

The grouping can be characterized by the phase grouping coefficient, $C_{PG}$, equal numerically to the ratio of azimuthal phase size of a "spoke" with emitted electrons to phase size of the "spoke" top close to the anode. Computed trajectories of motion of charges in "spokes" in a 2.45 GHz, 8-vanes magnetron at rotation of a "spoke" by $\pi$ with the ratio of the synchronous wave radial electric field to the static electric field $\varepsilon \approx 0.3$, (both taken on cathode) show that $C_{PG} \approx 2$, Fig. 1 [9].

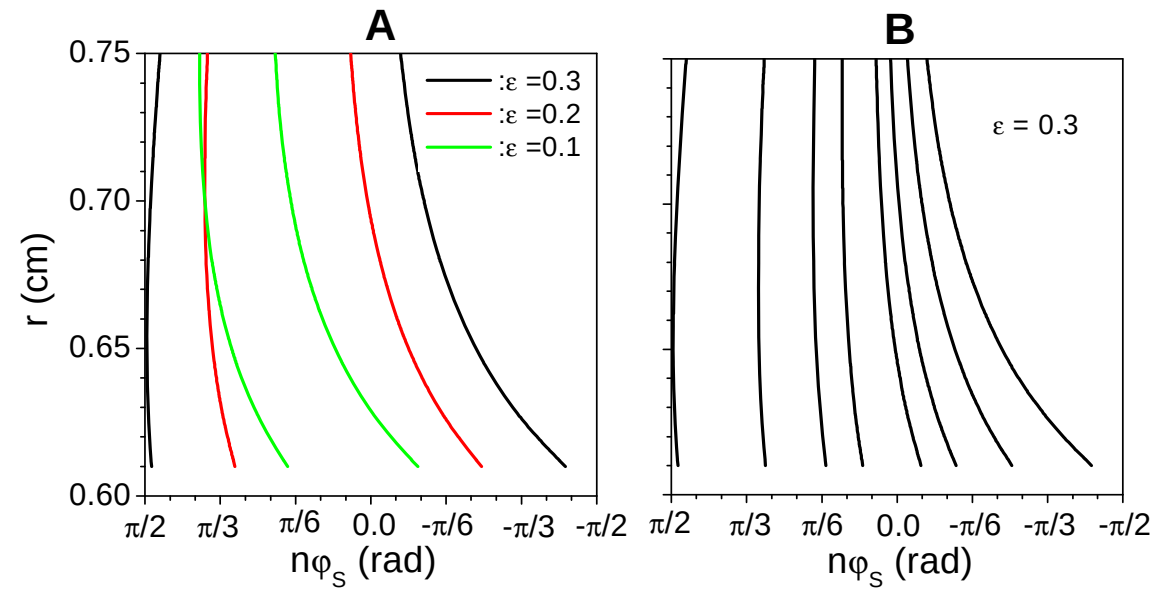

Fig. 1. Phase grouping of the charge drifting towards the magnetron anode in the magnetron model considered in [9]. Plot A shows a "spoke" boundaries vs. $\varepsilon$. Plot B shows trajectories of charges in a "spoke at $\varepsilon$=0.3.

The "spokes" delivering the electrons onto the anode are in the phase interval $(-\pi/2, \pi/2)$ out of $2\pi$ [9, 11]. The charges in a "spoke" producing coherent oscillations may reach the anode during of about 1-2 azimuthal rotations of a "spoke" around the cathode. One turn takes time of $2\pi \cdot n/\omega \sim 1.6$ ns in a 2.45 GHz 8 vanes magnetron. This time is much shorter than filling time of the magnetron cavity. Each subsequent "spoke" induces an increasing magnitude of the synchronous wave until saturation is reached, determined by the anode voltage. An increase of the phase grouping increases fields of the synchronous wave that, in turn, increases the phase grouping and so on. It means that the phase grouping in a magnetron is a fast, non-stationary, avalanche-like process.

Thus, one can seek an estimate of the build-up of a synchronous wave in such a process in the form of an exponential dependence on $C_{PG}$.

Since the azimuthal size of the phase-grouped charges close to the vanes slits is much less than the wavelength $\lambda$, generated by a magnetron, the induced injection-locked radiation of the phase-grouped Larmor electrons in "spokes" results in a coherent enhancement (gain) $G_C$, of its power. The coherent enhancement greatly increases the synchronous wave magnitude. This causes better phase-locking of the spontaneous oscillations and better phase grouping of charges. The processes lead to an exponential decrease of power of the phase-locked spontaneous oscillations, which in a working magnetron are converted into coherent and incoherent RF radiation.

Then, implying that the coherent radiation power is in the squared dependence on the radiating charges value and taking into account an exponential dependence of the avalanche-like process of the phase grouping, one can estimate the $G_C$ value as:

$$d(\lg G_C) \sim 2 \cdot I \cdot C_{PG} dI \, . \qquad (2)$$

Here $I$ is the magnetron anode current, $dI = I \cdot C_{PG}$ is the phase-grouped current inducing the synchronous wave. The product $I \cdot C_{PG}$ is dimensionless.

Thus, the phase-grouped Larmor electrons produce the coherently enhanced, injection-locked radiation (until the saturation is reached) with power of $P_C$:

$$P_C \sim P_S \cdot \exp(2 \cdot I^2 \cdot C_{PG}^2). \qquad (3)$$

Here: $P_S$ is power of the phase-locked spontaneous oscillations before the phase grouping of moving charges.

For magnetrons the number of Larmor electrons in the phase intervals passing towards the anode is ≤50% of the emitted one [11]. Table 1 shows $G_C$=$P_C$/$P_S$ estimates vs. $C_{PG}$ for 2.45 GHz microwave oven magnetron with the anode current $I$ ≈0.4 A. A droop of spontaneous radiation due to conversion in a coherent one and losses of coherency are not considered here.

Table 1. Estimations of the coherent enhancement $G_C$, vs. $C_{PG}$ in an operating microwave oven magnetron.

| $C_{PG}$ | 2 | 3 | 4 | 5 |
|---|---|---|---|---|
| $G_C$ | 3.6 | 17.8 | 170 | $3{\cdot}10^3$ |

The coherent generation of injection-locked magnetron as a process of the collective interaction of the phase-grouped electrons with a self-consistent field of the synchronous wave resembles process of coherent/quasi-coherent generation in a seeded FEL without an optical cavity [14].

## ON COHERENT GENERATION OF INJECTION-LOCKED MAGNETRONS

We have studied impact of injection-locking signal on the spontaneous and coherent radiations of 2.45 GHz, 1.2 kW magnetron type 2M137-IL in CW mode fed below the self-excitation threshold [8]. Figure 2 shows the carrier frequency offset measured with spectrum analyzer type E4445A at the resolution bandwidth of 100 kHz and the magnetron output power $P_{Mag}$ =100 W vs. various powers of injection-locking signal, $P_{Lock}$.

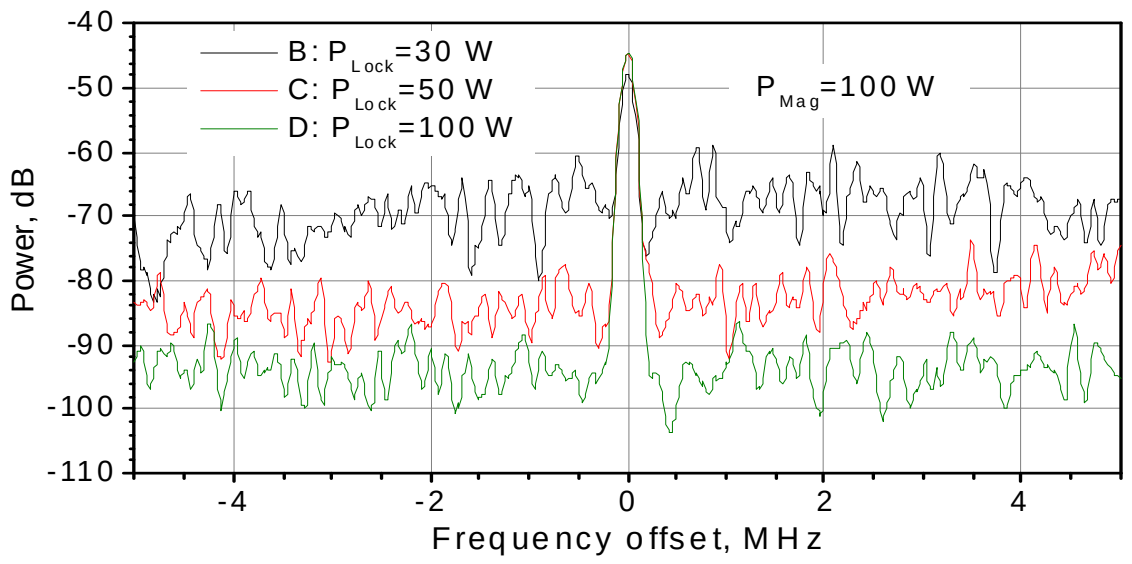

Fig. 2. Offset of the carrier frequency of magnetron type 2M137-IL measured at the magnetron output power of 100 W vs. power of the injection-locking signal, $P_{Lock}$.

Smoothing the traces one can estimate the spontaneous radiation (incoherent) power vs. power of the injection-locking signal close to the magnetron launching point, Fig. 3.

The trace at $P_{Lock}$ =30 W indicates the wide-band radiation with power ~1 W with the bandwidth of ≈7 MHz at the level of -3 dB. One can imply that this trace shows the phase-locked incoherent spontaneous radiation close to point of launching. Other traces show the incoherent spontaneous radiation greatly reduced due to conversion into the coherent magnetron RF generation.

A noticeable frequency shift of the phase-locked spontaneous oscillations at $P_{Lock}$ =50 W indicates frequency pushing of the oscillations caused by the anode current of launched magnetron. A further increase of the locking signal decreases intensity of the spontaneous oscillations converting them mainly into coherent these.

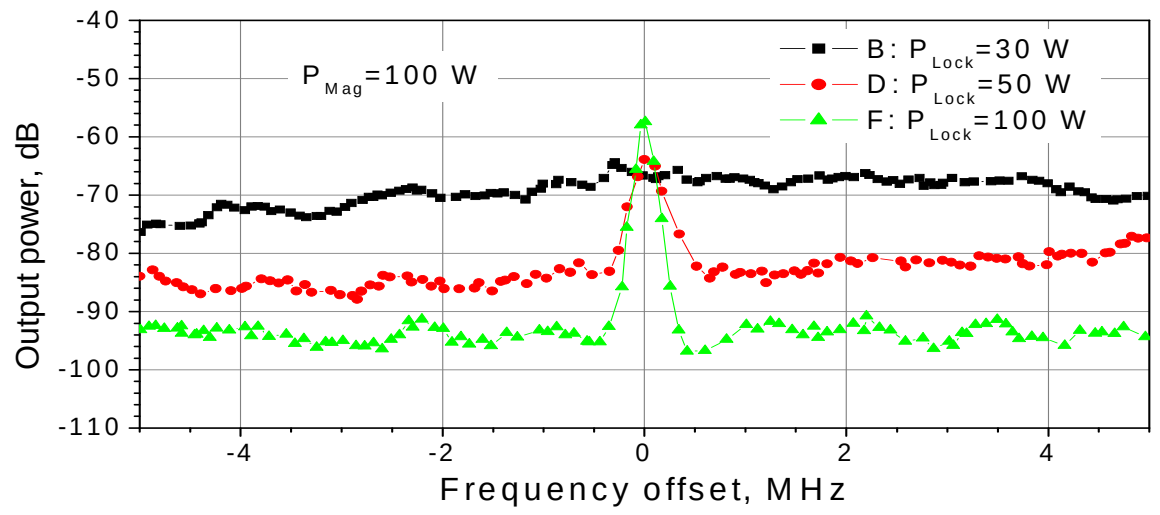

Fig. 3. Smoothed offset of the carrier frequency of the 2M137-IL magnetron at the magnetron output power of 100 W vs. power of the injection-locking signal, $P_{Lock}$.

Fit of the smoothed power of spontaneous incoherent radiation for various $P_{Lock}$ by an exponential decay, Fig. 4, shows an exponential droop of the spontaneous radiation due to a grow of the locking signal increasing magnitude of the synchronous wave.

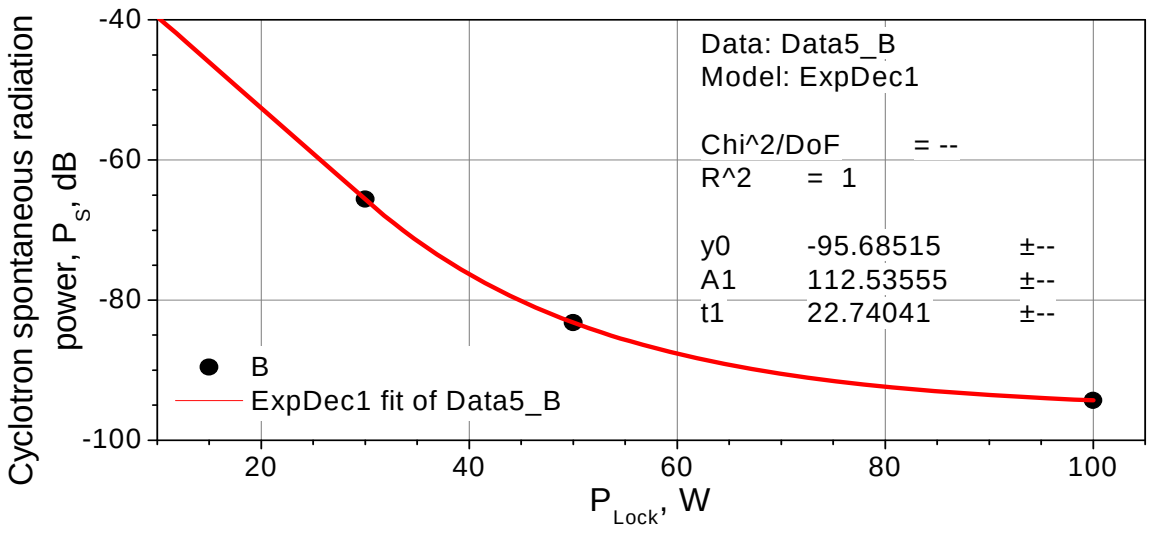

Fig. 4. Fit of power of the smoothed spontaneous radiation, shown by dots, by an exponential decay vs. power of the injection-locking signal $P_{Lock}$.

As noted above, this makes the spontaneous radiation mostly phase-locked and coherent due to the phase grouping. Its residual power characterizes the loss of coherency at the phase grouping vs. $P_{Lock}$. In the first approximation the loss of coherency is inversely proportional to the $G_C$ value.

Measuring the carrier frequency offset with resolution bandwidth of 5 Hz and large injection-locking signal we observed very narrow width of carrier frequency line $\Delta f_C$, ($\Delta f_C/f_C \leq 10^{-9}$), Fig. 5 [8, 9], with high spectral density.

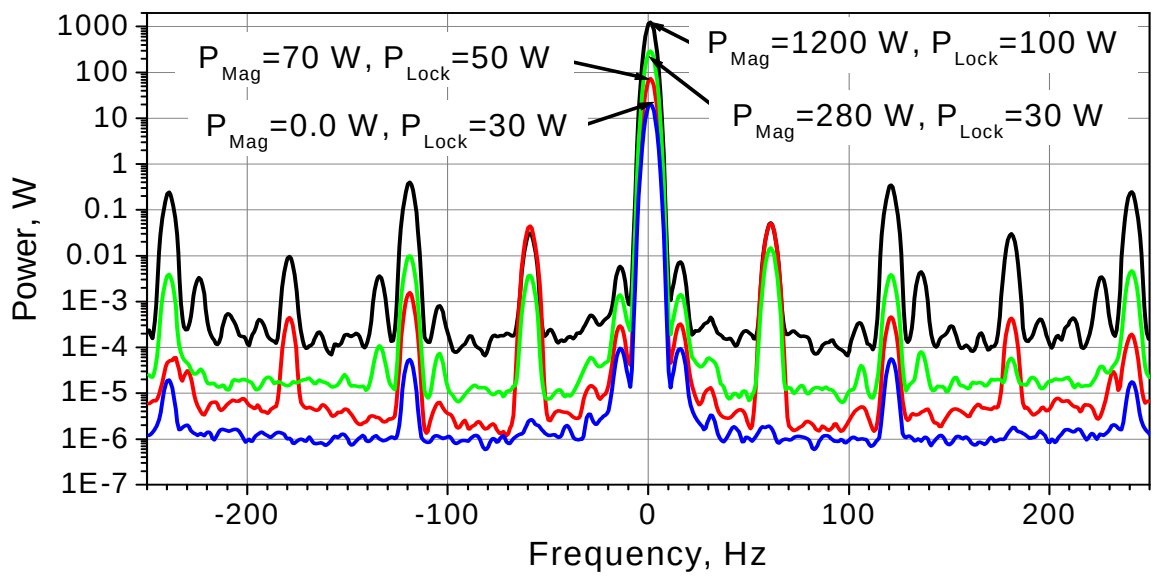

Fig. 5. Offset of the carrier frequency of magnetron type 2M137-IL (operating in CW mode) at various power levels of magnetron output, $P_{Mag}$, and the injection-locking signal, $P_{Lock}$.

The trace $P_{Mag}$ =0.0 W, $P_{Lock}$ =30 W shows the frequency offset of the injection-locking signal when the magnetron anode voltage is OFF. The trace $P_{Mag}$ =70 W, $P_{Lock}$ =50 W shows operation in so-called stimulated coherent generation mode.

Plots in Fig. 5 indicate that the magnetrons injection-locked by the signal providing a sufficient phase grouping operate almost as coherent oscillators greatly magnifying an injection-locked spontaneous radiation due to the coherent enhancement. Note that magnetrons operating without an injection-locking signal are also quasi-coherent oscillators. However, since the phase grouping is faster than the filling of the magnetron resonant system, the oscillations are starting on an instantaneous frequency of spontaneous oscillations, which is affected by ripples of the anode voltage, etc. This causes frequency instability of the "free running" magnetrons. Injection-locking of spontaneous oscillations eliminates this instability.

An additional decrease (≈25 dB) of a level of the measured spontaneous oscillations in Fig. 5 results from excluding numerous harmonics of power supplies due to the high resolution. The harmonics are caused by ripples of power supplies of master oscillator, TWT (used to amplify the injection-locking signal), the magnetron itself and alternating filament current of the tube [6].

Since the magnetron output power is determined by the coherent enhancement or its saturation, while the loss of coherency, reducing the coherent enhancement at a quite large injection-locking signal in the first approximation is proportional to the residual spontaneous radiated power, Fig. 3, one can assume that the higher value of ratio of the magnetron coherent power $P_{Mag}$, to the power of the residual spontaneous radiation $P_{RS}$, reflects better phase grouping i.e., the lower loss of coherency and the higher coherent enhancement. Thus, higher value of the ratio indicates magnetron parameters enabling better stability, higher efficiency and lover noise, i.e., parameters more suitable for SRF accelerators. High resolution enables better sensitivity of measurements of the ratio and allows finding better magnetron regimes for SRF accelerators.

Evaluations of the ratios of the carrier frequency power to the residual power of spontaneous radiation excluding sidebands, Fig. 5, are shown in Table 2. The incoherent injection-locked spontaneous radiation may contribute to the carrier frequency peak, but the contribution is negligibly low.

Table 2. Evaluations of the ratio $P_{Mag}/P_{RS}$,for magnetron 2M137-IL operating in CW mode for various measured values of magnetron anode voltage $U_{Mag}$, magnetron output and locking powers, $P_{Mag}$ and $P_{Lock}$, respectively, and the residual spontaneous radiation, $P_{RS}$.

| $U_{Mag}$, kV | $P_{Mag}$, W | $P_{Lock}$, W | $P_{RS}$, W | $P_{Mag}/P_{RS}$ |
|---|---|---|---|---|
| 3.90 | 70 | 50 | $1.9\cdot10^{-6}$ | $4.1\cdot10^{7}$ |
| 4.01 | 280 | 30 | $1.0\cdot10^{-5}$ | $2.2\cdot10^{7}$ |
| 4.09 | 1200 | 100 | $1.5\cdot10^{-4}$ | $8.6\cdot10^{6}$ |

The coherent power, shown in Table 1 and 2, evaluated by eq. (3) for $C_{PG}$ ≈3.6, using the estimate of measured phase-locked spontaneous radiation power of ~1 W and measured for stimulated coherent generation mode, respectively, are in agreement. For this mode the ratio $P_{Mag}/P_{RS}$ is largest that indicates the best phase grouping in the stimulated coherent generation mode. Modeling of operation of a 10-vanes microwave oven magnetron (without an injection-locking signal) using the PIC code [15] shows that $C_{PG}$ ≈3.8. Thus, the increased number of vanes improves phase grouping, which increases the efficiency of the magnetron [10].

Note that earlier attempt of measurements with 1 kW L-band magnetron being injection-locked by about of -10 dB signal and fed below the self-excitation threshold voltage showed "…spurious oscillations at -50 dB to -60 dB level… at frequencies which were not the oscillation frequencies of the magnetron…" [16]. Most likely this indicates a spontaneous phase-locked radiation, which the author then could not recognize.

In agreement with the kinetic model of magnetron operation [9], the performed measurements indicate that an increase of the injection-locking signal at some decrease of the magnetron anode voltage magnifies the coherent enhancement, means the phase grouping and reduces fluctuation of the coherent enhancement due to more stable increment of magnitude of the synchronous wave. Finally this stabilizes the coherent enhancement, reduces noise and reduces losses of coherency, i.e., anode losses [11]. This may also reduce loss of electrons from "spokes" maximizing the magnetron efficiency [8].

Thus, parameterization of the phase grouping by coefficient $C_{PG}$, and the coherent enhancement value $G_C$, depending on it, indicates how to improve the magnetron design, e.g., increasing number of vanes [10], and how to choose parameters of operation and control, e.g., using the mode of stimulated coherent generation for higher stability of the carrier frequency at very high its spectral density, higher efficiency and best controllability of the magnetron RF source that is most suitable for SRF accelerators.

Note, that a notable decrease of the magnetron anode voltage when the tube operates in stimulated coherent generation mode, Fig. 6, [10], increases the tube reliability and is a promising way to increase its lifetime.

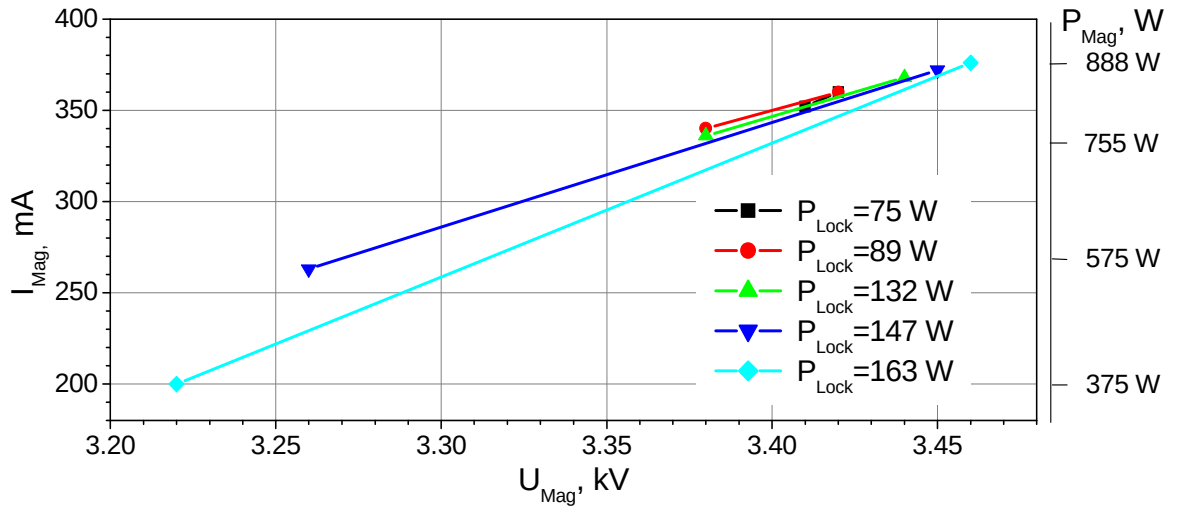

Fig. 6. Measured ranges of anode voltage, $U_{Mag}$, and magnetron current, $I_{Mag}$, in the stimulated coherent generation mode for 2.45 GHz, 945 W magnetron type 2M219G (having the self-excitation threshold voltage of 3.69 kV) at various power levels of the injection-locking signal $P_{Lock}$. The right scale shows measured RF power of the magnetron, $P_{Mag}$, vs. the magnetron current, $I_{Mag}$.

The plots shown in Fig. 6 demonstrate that coherent stimulated generation mode of a magnetron enables operation at quite wide range of current (power) control being most efficient comparing to other RF sources. Performance of the magnetron type 2M219G in stimulated coherent generation mode compared to traditionally used regime demonstrates the increase of magnetron efficiency by more than 10% [10].

In pulse operation in stimulated coherent generation mode the magnetrons provide 100% pulse modulation of the output signal due to 100% pulse modulation of the synchronous wave without modulation of the cathode voltage, if the tube with the anode voltage respective to values shown in Fig. 6 is driven by a pulse (gated) RF injection-locking signal [10].

This enables a great reducing of the capital cost of pulsed RF sources for pulse SRF accelerators and an addition increase of the source's efficiency due to absence of pulse HV modulators causing additional losses of power.

Note that utilization of two-cascade magnetrons [9], enables reducing the injection-locking signal power necessary for stimulated coherent generation mode by ≈10 dB making the two-cascade magnetron transmitters quite affordable.

## IMPACT OF THE INJECTION-LOCKING SIGNAL ON A BANDWIDTH OF A MAGNETRON CONTROL

We studied impact of strength of the injection-locking signal on the bandwidth of the magnetron control measuring transfer function magnitude characteristic in phase modulation domain [5], and measuring transfer function phase characteristic [9]. The experiments were performed with 2.45 MHz, 1 kW magnetrons operating in pulse mode at the power supply pulse duration of 5 ms, using phase modulation of the injection-locking signal.
The admissible bandwidth of the phase and power control in magnetrons is determined by stability of the LLRF system with feedback loops at the dynamic control. The bandwidth of control was determined by measured transfer function characteristics for levels of the transfer function magnitude and phase characteristics of -3 dB and 45 deg., respectively, assuming first order filters in the LLRF system. Fig. 7 shows the admissible bandwidth of control, $BW_C$, of 2.45 GHz, 1 kW microwave oven magnetrons determined vs. the locking signal power.

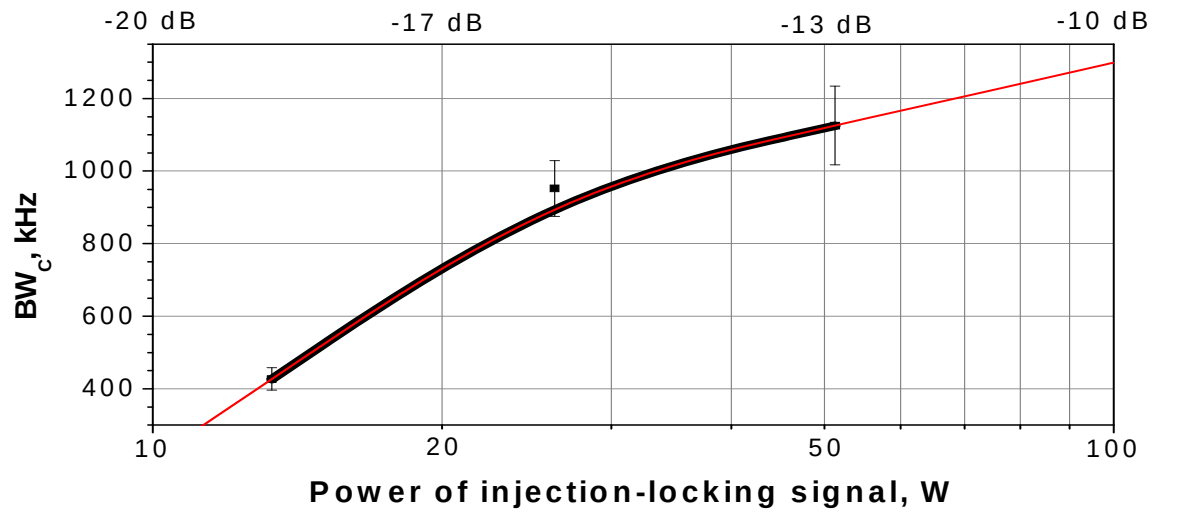

Fig. 7. The admissible bandwidth of control of 2.45 GHz microwave oven magnetrons determined by measured transfer function characteristics. Black bold line shows the range and results of measurements, red lines show extrapolation with B-spline fit.

Plots in Fig. 7 show that the locking signals of -15 dB and larger increase the control bandwidth approximately logarithmically. For first order filters the out-of-band roll-off is 20 dB/decade. For L-band magnetrons intended for modern SRF accelerators one can expect the bandwidth of control of about 100-500 kHz. This will allow attenuation of amplitude of parasitic modulation caused by "microphonics", etc., significantly more than by 60 dB.

Thus, an increase of the locking signal amplifies phase grouping and increases the magnetron control bandwidth.

## SUMMARY

We have developed and experimentally proved parameterization of the phase grouping which is a useful tool for analysis of operation and control of magnetrons enabling to find regimes with higher efficiency, lower phase noise and widest bandwidth of the phase and power control. The developed technique clarifies impact of phase grouping of Larmor electrons in "spokes" on the magnetron operation and control by a simple representation and estimates of the phase grouping process. Simulation of operation of a magnetron using the analytical (kinetic) model and the parameterization enable to improve the magnetron design. This will allow usage of required regimes and improvements in affordable magnetron transmitters to make them most efficient and suitable for feeding SRF cavities in various superconducting accelerator projects. Presented estimates compared to experiments indicate that the amplified phase grouping in the stimulated coherent generation mode magnifies the coherent enhancement increasing the magnetron efficiency and reducing fluctuations of the self-consistent field of the synchronous wave that increases the magnetron stability and decreases noise. In this mode a magnetron operates as an almost-coherent oscillator (at low loss of coherency) with highest spectral density of the carrier frequency. A sufficient signal injection-locking the magnetron resulting in improved phase grouping of electrons reduces loss of the electrons delivered by "spokes" towards the anode. This will reduce the electron back stream and should decrease somewhat the magnetron cathode temperature. The magnetron anode voltage, reduced when the tube operates in the stimulated coherent generation mode, indicates that use of this mode is a promising way to increase the magnetron reliability and longevity.

## ACKNOWLEDGEMENT

The work was supported by collaboration of Muons, Inc and Fermi National Accelerator Laboratory, Fermi Research Alliance, LLC under CRADA-FRA-2017-0026.

We are very thankful to Dr. Ya. Derbenev and Dr. V. Balbekov for fruitful discussions, and Dr. V. Lebedev for stimulating discussions and help in the paper editing.